\documentclass{elsart}
\usepackage{epsfig}
\usepackage{graphicx}
\begin{document}
\begin{frontmatter}
\title{Modulational instability of matter waves under
strong nonlinearity management }
\author{F.Kh. Abdullaev\corauthref{cor1}},
\author{A.A. Abdumalikov},
\author{R.M. Galimzyanov}
\address{Physical-Technical Institute of the Uzbek Academy of Sciences,
2-b, G. Mavlyanov str., 100084, Tashkent, Uzbekistan }
\corauth[cor1]{Corresponding author. E-mail: fatkh@uzsci.net}
\begin{abstract}
We study modulational instability of matter-waves in Bose-Einstein
condensates (BEC) under strong temporal nonlinearity-management.
Both BEC in an optical lattice and homogeneous BEC are considered
in the framework of the Gross-Pitaevskii equation, averaged over
rapid time modulations. For a BEC in an optical lattice, it is
shown that the loop formed on a dispersion curve undergoes
transformation due to the nonlinearity-management. A critical
strength for the nonlinearity-management strength is obtained that
changes the character of instability of an attractive condensate.
MI is shown to occur below(above) the threshold for the
positive(negative) effective mass. The enhancement of number of
atoms in the nonlinearity-managed gap soliton is revealed.
\end{abstract}

\begin{keyword}
modulational instability, matter wave, Feshbach resonance
management, optical lattice, gap soliton  \PACS 03.75.Lm;
03.75.-b;30.Jp
\end{keyword}
\end{frontmatter}

\section{Introduction}

The phenomenon of modulational instability (MI) of nonlinear plane
waves under different types of management of the system parameters
has been the subject of intensive research over the last years
\cite{ADG}. Main emphasis was given to dispersion-management and
nonlinearity-management. In nonlinear optics strong and rapid
modulations of the fiber dispersion is achieved by periodic
arrangement of fiber spans with alternating sign of the
dispersion. Dispersion-managed solitons supported by such a system
have essential advantages over conventional optical solitons for
long distance communication purposes \cite{Doran,ADKL,BK}.
Modulations of the nonlinearity is a challenging problem also in
fiber ring lasers and in generation of Faraday waves in
Bose-Einstein condensates (BEC)
\cite{Abdullaev94,ADBS,Stal,Kevrekidis03,Engels}. MI in the form
of Faraday waves can be observed both in {\it attractive} and {\it
repulsive} condensates. Recent observation of the MI in optical
media resulted from the periodic modulation of the nonlinearity in
the evolution variable, confirms the existence of parametric
resonances in the MI growth rate \cite{ADBS,Kevrekidis}. Faraday
waves (parametrically excited waves) in a BEC emerging from
temporal periodic variation of the atomic scattering length have
been studied in \cite{Engels}. Such type of modulations can be
achieved by variation of the external magnetic field near Feshbach
resonances (FR). The corresponding technique is known as FR
management. In the Gross-Pitaevskii equation this corresponds to a
temporal variation of the mean-field nonlinearity, i.e.  to the
nonlinearity-management. MI in a harmonically trapped BEC under FR
management has been investigated in \cite{Rapti}.

Recently the strong dispersion-management has been applied to the
dynamics of nonlinear periodic waves, namely cnoidal waves, in
optical fibers \cite{Kartashov1,Korneev}. In these works the
existence of dispersion-managed cnoidal waves and strong deviation
of the stability borders of these waves from the ones of standard
cnoidal wave solutions of the nonlinear Schr\"odinger equation
(NLSE) have been established. Extension of the stability regions
of some types of nonlinear periodic waves  can be due to the
different scenarios for the onset of MI of the background plane
waves. Adiabatic FR management for cnoidal waves in optical
lattices has been considered in \cite{BK05,AKKB}. The case of
strong nonlinearity-management remains unexplored.

The strong nonlinearity-management may be an effective tool for
stabilization of matter-wave solitons in multi-dimensional
attractive BEC
\cite{SU,ACMK,Mont,AG,Stefanov,KonPac,MPM,MRPM,Itin,Liu}. In the
context of nonlinear optics such stabilization mechanism was first
discussed in \cite{Berge,Towers}. The phenomenon of MI is
particularly important for generation of soliton trains in BEC
with controlled spatial arrangement (repetition rate). MI of BEC
in linear and nonlinear optical lattices in the absence of
time-periodic nonlinearity-management has been investigated in our
recent work \cite{AbdAbdGal}. Here we consider both the MI of a
homogeneous BEC and MI of a BEC loaded in an optical lattice under
FR management. The gap soliton structure existing in a BEC with
the zero background scattering length $(a_{sb}=0)$ has been
investigated in Ref.~\cite{Porter1}. The couple-mode theory can be
used to analyze MI of nonlinear plane waves in an optical lattice
subject to FR management. In our investigations particular
interest we will paid to the properties of loop structures
emerging in the band gaps (forbidden band).

In the present paper we investigate  nonlinear dispersion
relations and the process of MI in a BEC under strong temporal
nonlinearity management (SNM). The outline of the paper is as
follows. The mathematical model is formulated in Section 2. MI in
a homogeneous BEC under SNM is considered in Section 3. The
nonlinear dispersion relation and loop structures for BEC in an
optical lattice under SNM are analyzed in Section 4 using the
coupled-mode theory. This section also includes the regions of MI
found in different areas of the band structure. The properties of
gap solitons are investigated in  Section 5. Section 6 is devoted
to details of our numerical procedure. In the final Section 7 we
summarize our main results.

\section{The model}
Let us consider a BEC under  temporal Feshbach resonance
management when the  scattering length $a_{s}$ varies in time.
Then an elongated BEC can be described by the quasi-1D GP equation
with a periodic potential ( optical lattice) and the
time-dependent management of the coefficient of nonlinearity
\begin{equation}
i\hbar\psi_t = -\frac{\hbar^2}{2m}\psi_{xx} + V(x)\psi -
g_{1D}(t)|\psi|^2\psi, \label{GP}
\end{equation}
where $g_{1D}(t) = 2\hbar a_{s}(t)\omega_{\perp}$ is the mean
field nonlinearity coefficient, $\
\omega_{\perp}$ is the transverse oscillator frequency and $V(x)=
V_0\cos^2(kx)$ is an optical lattice potential,
$\int_{-\infty}^{\infty} dx |\psi|^2 =N$, $N$ is the number of
atoms. In dimensionless units we have
$$
x \rightarrow kx, \ t \rightarrow \omega_R t, \  \epsilon =
\frac{V_0}{2E_R}, \ E_R = \frac{\hbar^2 k^2}{2m},\
\omega_R=E_R/\hbar, \ u = \sqrt{\frac{2\hbar a_s
\omega_{\perp}}{E_R}}\psi e^{-i\epsilon t}.
$$ Eq.~(\ref{GP}) takes the form of the NLSE with varying in
time mean field nonlinearity coefficient
\begin{equation}\label{nls}
iu_t + u_{xx} + \gamma(t)|u|^2 u   - 2\epsilon\cos(2x)u = 0,
\end{equation}
where $\gamma(t)$ describes the strong nonlinearity-management and
has the form
\begin{equation}\label{gamma}
\gamma(t) = \gamma_{0} +
\frac{1}{\mu}\gamma_{1}\left(\frac{t}{\mu}\right), \
\int_{0}^{1}\gamma_{1}(\tau)d\tau = 0, \ \tau = \frac{t}{\mu},\
\mu \ll 1.
\end{equation}

 This model has been considered in recent papers
\cite{BK05,AKKB,Porter1}. Specifically, in  works \cite{BK05,AKKB}
the evolution of nonlinear periodic waves under adiabatic
time-variation of the scattering length has been studied and a
possibility of generation of a train of solitons by such a
management scheme has been shown. Properties of gap solitons under
the strong management of nonlinearity were analyzed based on the
coupled mode system of equations in \cite{Porter1}. In this work
the gap soliton solutions and their stability for the case
$\gamma_0 =0$ were investigated. Here we will study MI of
nonlinear plane waves in a BEC (without and with an optical
lattice) under SNM, as well as properties of gap solitons in the
model (\ref{nls}) for nonzero value of $\gamma_0$. In particular
we will analyze the possibility of enhancement of number of atoms
in the gap soliton under SNM.

In deriving averaged equation we follow the works
\cite{Porter1,Zhar} and use the transformation
\begin{equation}\label{trans}
u(x,t) = e^{i\gamma_{-1}(t)|v|^2}v(x,t),\ \  \gamma_{-1}(\tau) =
\int_{0}^{1}\gamma(\tau^{\prime}) d\tau^{\prime}-
\int_0^1\int_0^{\tau}\gamma(\tau^{\prime})d\tau^{\prime}d\tau.
\end{equation}
Supposing the parameter $\mu$ to be small (that corresponds to
high frequencies of modulation) unknown function $v$ can be
expanded in series as
\begin{equation}\label{exp}
v = w + \mu v_1 + \mu^2 v_2 +... ,
\end{equation}
where unknown $w$ is a slowly varying function. Using
transformation (\ref{trans}) and expansion (\ref{exp}) in
governing equation (\ref{nls}) with posterior averaging over the
period of rapid modulation, we arrive at the following averaged
equation for $w$ \cite{Zhar}
\begin{eqnarray}\label{nm_nls}
iw_t &+& w_{xx} + \gamma_{0}|w|^2 w - 2\epsilon\cos(2x)w  +
\sigma^2 [ 2(|w|^2)_{xx}|w|^2  +\nonumber\\ &&((|w|^2)_{x})^2]w
=0.
\end{eqnarray}
Parameter $\sigma$ is defined as $\sigma^2 = \int_0^1
\gamma_{-1}^2 d\tau $. For particular case of sinusoidal
modulations $\gamma_{1} = h\sin(\omega t)$ we have $\sigma^2 =
h^2/(2\omega^2) \sim O(1)$ $( \omega = 1/\mu)$. For the step-like
modulation with the same amplitude $h$ and frequency $\omega$ we
have $\sigma^2 = h^2/\omega^2.$

This form of averaged equation can be also obtained for the case
of the weak nonlinearity management when $\gamma = \gamma_0 +
\gamma_1(t/\mu)$, with $\sigma^2\ll 1$ \cite{Zhar,ATMK}.

\section{Modulational instability of nonlinear plane wave in a homogeneous media}
Now let us consider the case when the optical lattice is switched
off, i.e. $\epsilon = 0$ in Eq.~(\ref{nls}).  The MI of a
nonlinear plane wave $w = A\exp(i(\gamma_{0} A^2 t)$ can be
explored using the linear stability analysis, i.e. looking for the
solution in the form
\begin{equation}\label{perpw}
w = (A+\psi(x,t))\exp[i\gamma_{0} A^2 t ], \ \psi \ll A.
\end{equation}
We have the following equation for $\psi$
\begin{equation}
i\psi_t + \psi_{xx} + \gamma_0 A^2 (\psi + \psi^{\ast}) +
2\sigma^2 A^4 (\psi_{xx} + \psi_{xx}^{\ast}) = 0.
\end{equation}
Representing $\psi = \psi_r + i\psi_i$ and performing Fourier
transformation $\psi_r(\psi_i)(x,t) = \int dk
\bar{u}(\bar{v})(k,t)\exp(ikx)$
 we get  the
dispersion relation
\begin{equation}
p^2 = k^2  [2\gamma_0 A^2-(1 + 4\sigma^2 A^4)k^2 ].
\end{equation}
Instability region corresponds to the condition
$p^2 > 0$. Thus we obtain
\begin{equation}\label{MI}
k^2 \leq \frac{2\gamma_0 A^2 }{1 + 4\sigma^2 A^4}.
\end{equation}
The maximum of the MI gain is achieved at the value of the wave
number
\begin{equation}\label{max}
k_c = \sqrt{\frac{\gamma_0 }{1 + 4\sigma^2 A^4}}A.
\end{equation}
Maximal value of the MI growth rate is
\begin{equation}\label{gain1}
p_c = \frac{\gamma_0 A^2}{\sqrt{1 + 4\sigma^2 A^4}}.
\end{equation}
Thus we find that under the temporal nonlinearity management the
MI growth rate is decreased by a factor of $\sqrt{1 + 4\sigma^2
A^4}$. Such decrease of the gain is due to the defocusing effect
induced by the nonlinearity  management. This observation explains
the stabilizing role of the strong nonlinearity management in a
higher dimensional attractive BEC \cite{SU,ACMK,Adhikari,Saito2}.

\begin{figure}[htbp]
\vspace{0.5cm}
\begin{center}
\includegraphics[width=6cm,angle=-90]{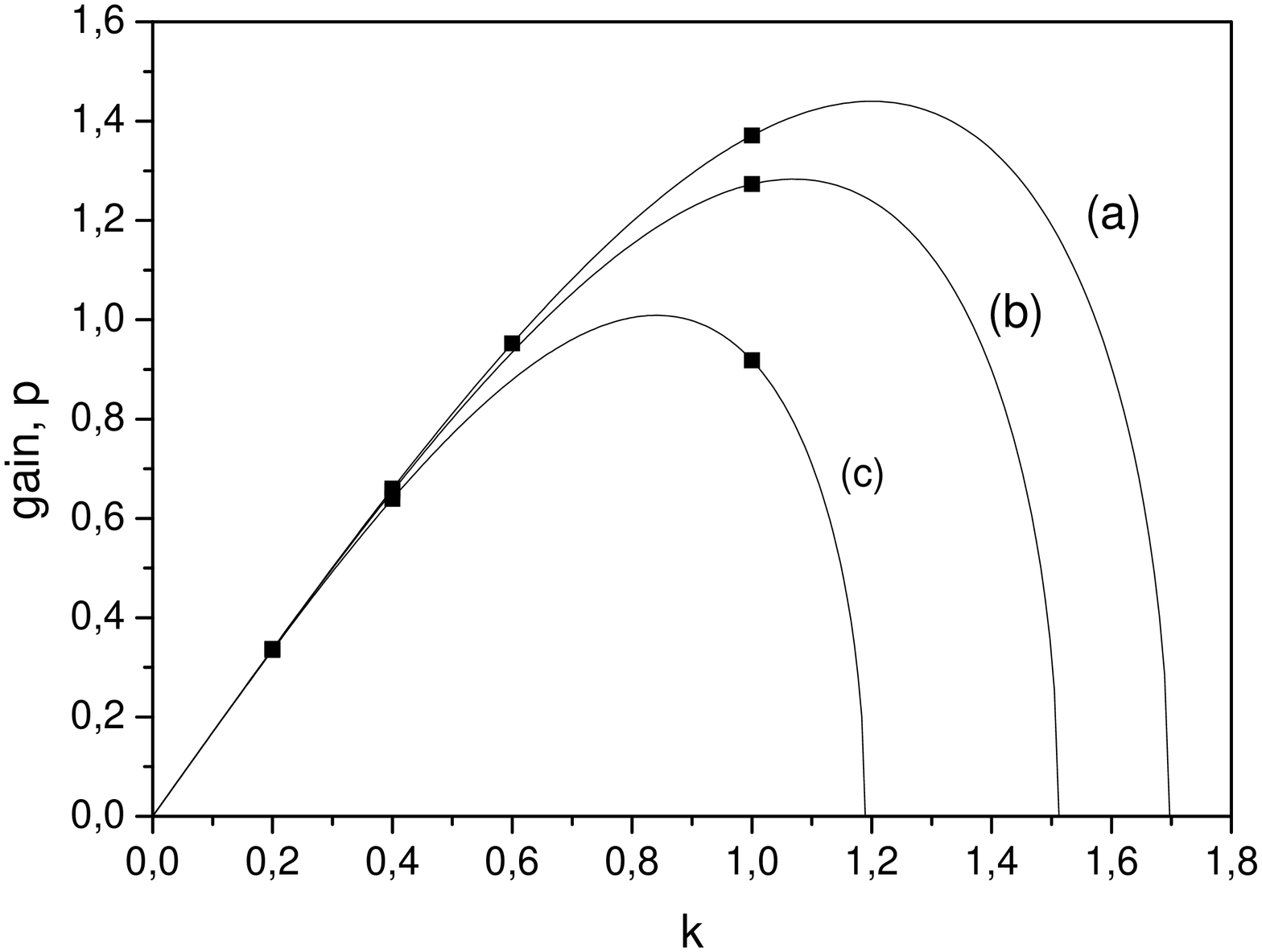}
\end{center}
\caption{MI gain $p$ versus the wave number modulations $k$. Three
curves  correspond to the cases when: (a) nonlinear management is
turned off, $\sigma^2 = 0$; (b), (c)  nonlinear management is
turned on, with $A = 1.2$, $\sigma^2 = 0.125$ and $\sigma^2 =
0.5$. Filled squares correspond to gains obtained from full PDE
simulations.} \label{gain}
\end{figure}

\begin{figure}[htbp]
\begin{center}
\includegraphics[width=5cm,angle=-90]{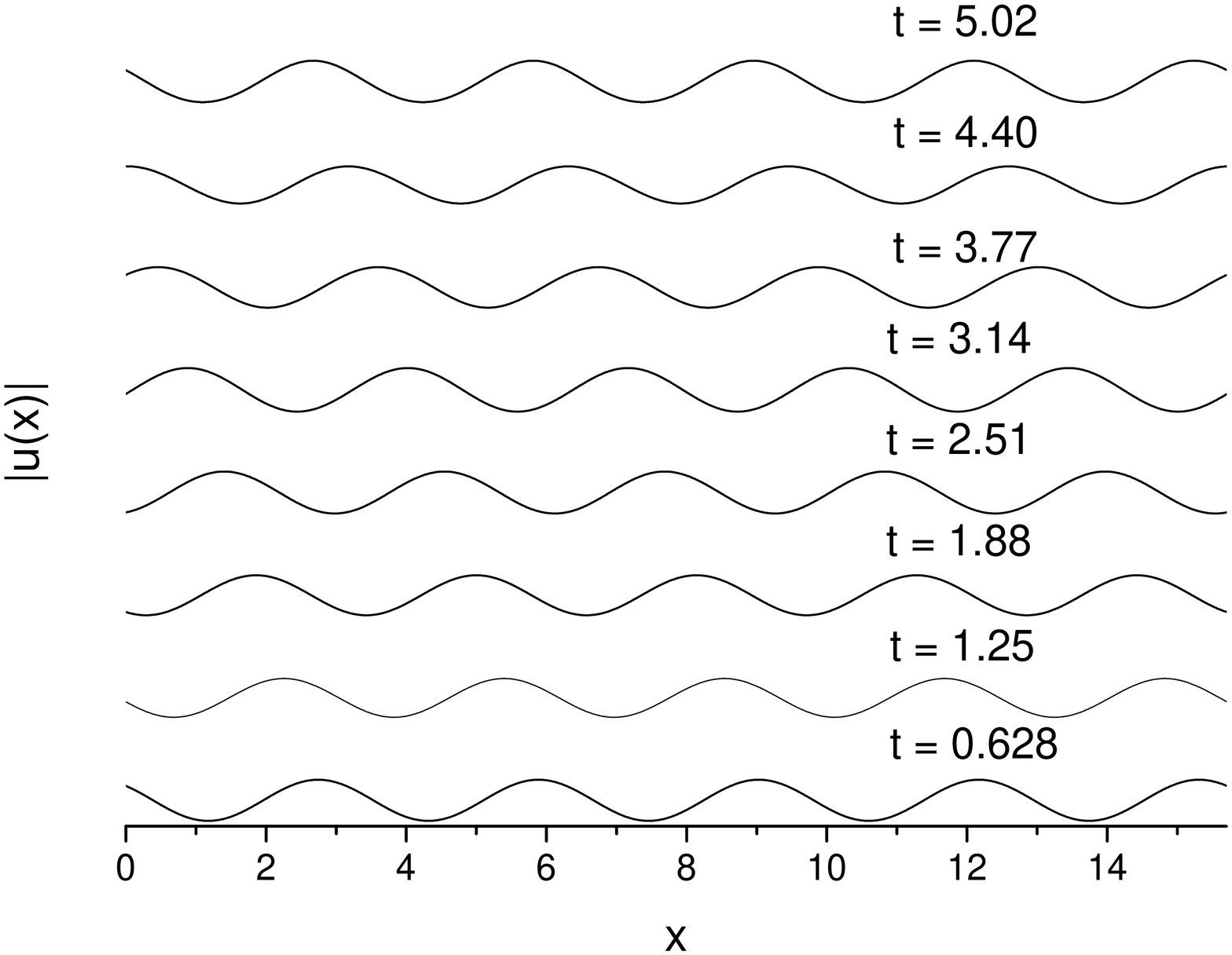}
\end{center} \hspace{1.5cm}
\caption{ Evolution of the small spatially periodic perturbation
when $p$ and the wave number $k$ are in the region of stability
and inequality (\ref{MI}) is not fulfilled.  The case with
$\sigma^2 = 0.02, \ k=2, A = 1.2$ is presented.} \label{stab}
\end{figure}

\begin{figure}[htbp]
\includegraphics[width=5cm,angle=-90]{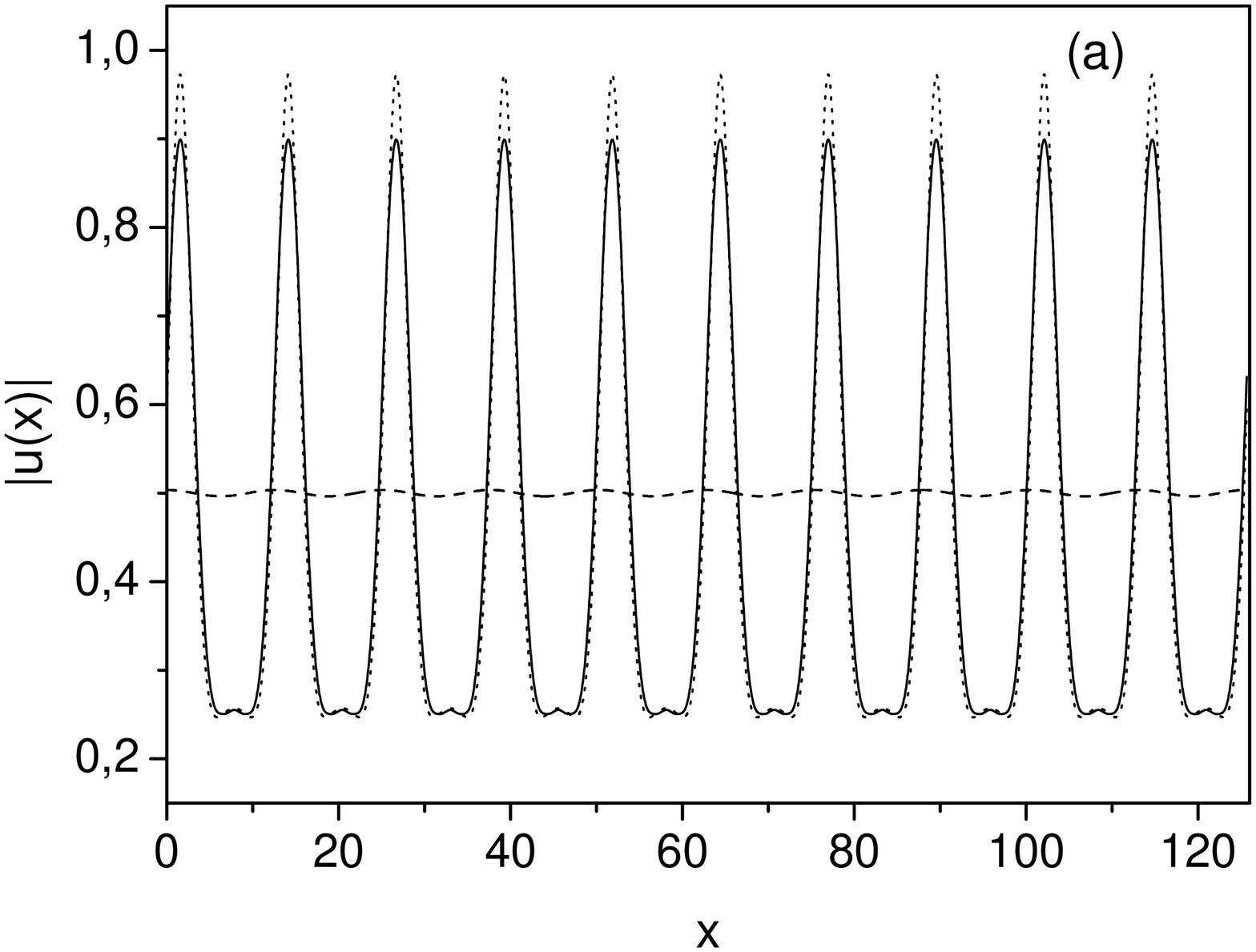}
\hspace{0.5cm}
\includegraphics[width=5cm,angle=-90]{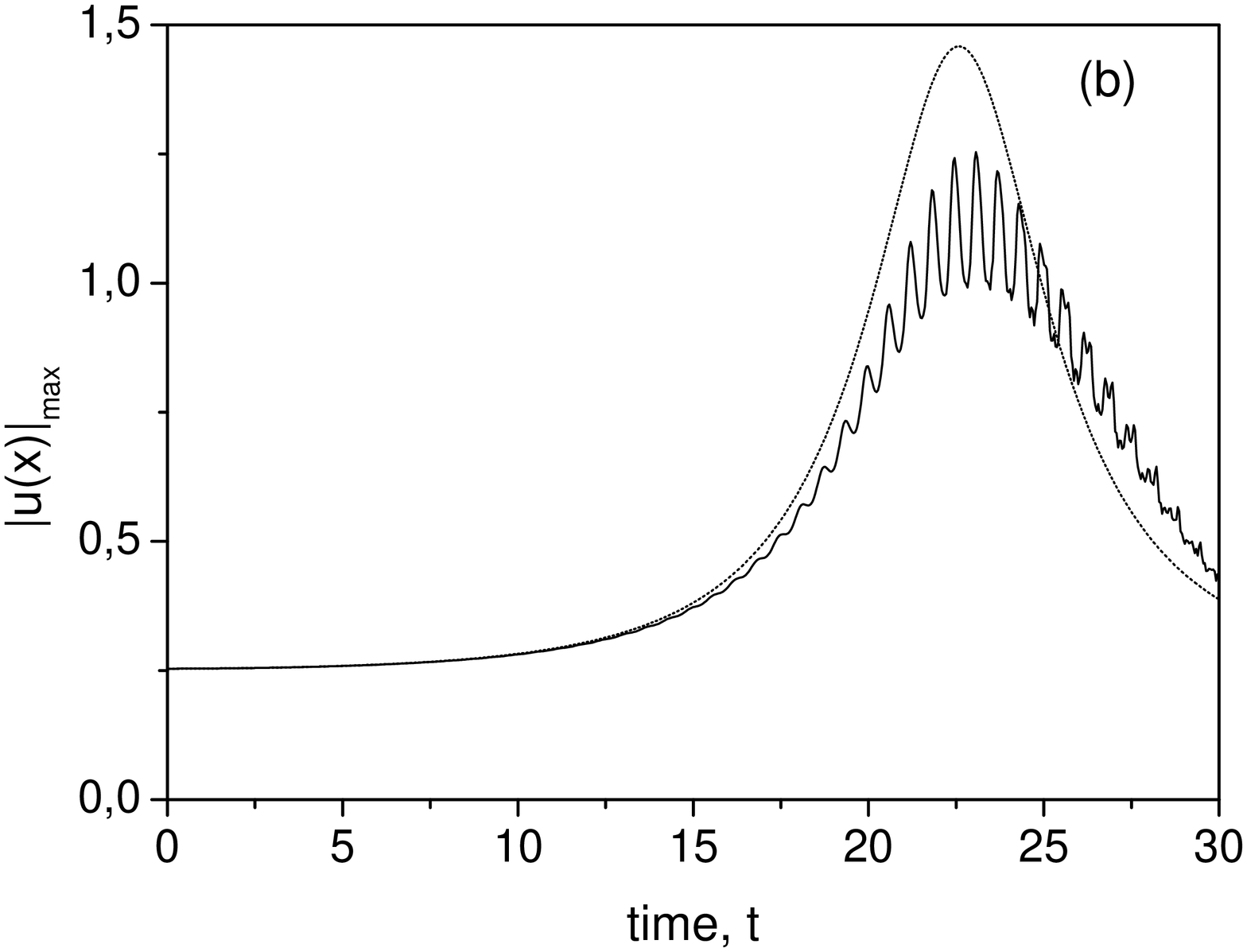}
\hspace{1.5cm} \caption{ Development of small spatially periodic
perturbations into a soliton train when parameters are in the
region of instability. Plot (a) depicts the field profiles
$|u(x)|$ at different times. Dashed line stands for initial small
modulations at $t = 0$, dotted (solid) line is for the case, when
SNM is turned off (on) at $t = 20.1$. Plot (b) depicts time
evolution of the maximal value of $|u(x)|$. Dotted (solid) line
correspond to turned off (on) SNM. The parameters are $k = 0.5, \
A = 0.5, \ \sigma^2 = 0$ ($\sigma^2 = 0.125$).} \label{instab}
\end{figure}

Numerical simulations of the 1D GP equation (\ref{nls}) with a
strong nonlinearity management confirm these predictions. In
Fig.~\ref{gain} we plot the MI gain versus the wave number of
modulations $k$ for three different cases with $\gamma_0 = 1$ and
$\omega = 10$: (a) when the nonlinearity-management is absent,
$\sigma^2 = 0$ and when the management is present (b) $\sigma^2 =
0.125 (h = 5)$, (c) $\sigma^2 = 0.5 (h = 10)$. One  can observe a
good agreement between the theory and numerical simulations for
the value and the position of the MI gain maximum given by Eqs.
(\ref{max}) and (\ref{gain1}).  In Fig.~\ref{stab} we plot the
profiles of the field module $|u(x)|$ in the region of stability.
Fig.~\ref{instab} depicts the case of breakdown of the stability
caused by increasing the strength of the nonlinearity-management,
$\sigma^2$.
One can see that modulation in an initial plane wave evolves into
a train of solitons when the wave number of the modulation is in
the region of instability. As can be seen from
Fig.~\ref{instab}(a) even moderate nonlinearity management
($\sigma^2 = 0.125$) causes notable decreasing in the amplitude of
solitons.

\section{MI in a BEC loaded in an optical lattice and nonlinearity-management}
\subsection{Nonlinear dispersion relation. The loop structure.}
The analysis performed in the previous section was relevant to a
BEC without optical lattice potential. In the presence of an
optical lattice the band structure strongly affects the process of
MI \cite{Sterke,KS}. Equations of the coupled-mode theory for the
GP equation (\ref{nm_nls}) with shallow optical lattice have been
obtained in \cite{Porter1}. The wave function can be represented
in the form of superposition of backward and forward propagating
waves
\begin{equation}\label{pwloop}
w(x,t) = \sqrt{\epsilon}\left(A(X,T)e^{ix} +
B(X,T)e^{-ix}\right)e^{-it}.
\end{equation}
where $X = \epsilon x, \ T = \epsilon t$ are slow variables.
Substituting this into the averaged equation we get the following
coupled mode system of equations
\begin{eqnarray}\label{cm1}
iA_T + 2iA_{X} =  B - \gamma_{0}(|A|^2 + 2|B|^2)A +
8\epsilon\sigma^2(2|A|^2 + |B|^2) |B|^2 A,\\
\label{cm1a} iB_{T}-2iB_X =  A - \gamma_0 (2|A|^2 + |B|^2)B +
8\epsilon\sigma^2 (|A|^2 + 2|B|^2)|A|^2 B.
\end{eqnarray}
   In derivation of this system, the derivatives of the nonlinear
terms have been neglected as the terms of the next order of
smallness with respect to $\epsilon$. The group velocity varies in
the interval $-2 < v < 2$, in physical units that corresponds to
$-v_R < v < v_R, \ v_R = \hbar k/m.$ This system describes two
counter propagating waves, with the cubic self phase modulation
term and cubic and quintic cross-phase modulation terms. The
quintic cross modulation term describes effect of the Feshbach
resonance management.  Note that this system has a similarity with
the one previously considered for description of MI in the
cubic-quintic NLSE with the Bragg grating \cite{Porz}. However, as
distinct from that model, no self-phase modulation quintic terms
like $|A|^4 A$ and $|B|^4 B$ present in our model. The absence of
these terms changes significantly the MI process in NM systems in
comparison with the standart cubic-quintic NLS model.

The plane wave solutions of Eqs.(\ref{cm1}) and (\ref{cm1a}) are
looked for in the form
$$
    A =\frac{\alpha  }{\sqrt{1+f^2}}e^{i(QX - \Omega T)},\quad
    B =\frac{\alpha f}{\sqrt{1+f^2}}e^{i(QX - \Omega T)},
$$
where $\alpha = |A|^2 + |B|^2$. The parameter $f$ defines the
weight of the forward and backward propagating waves. The case
$|f|> 1$ corresponds to the domination of the backward wave.
Substituting these expressions into the system (\ref{cm1}) and
(\ref{cm1a}), we obtain nonlinear dispersion relation
\begin{eqnarray}\label{cm2}
\Omega &=& -\frac{3\gamma_0}{2}\alpha^2 +
\frac{1}{2}\frac{1+f^2}{f} +
\frac{4\epsilon\sigma^2\alpha^4}{(1+f^2)^2}(f^4 + 4f^2 + 1),\\
\label{cm3} Q &=& \frac{( 8\epsilon\sigma^2\alpha^2 -
\gamma_0)\alpha^2 }{4}\frac{1-f^2}{1+f^2} +
\frac{1}{4}\frac{1-f^2}{f}.
\end{eqnarray}
 The parameter $f$ determines the position on the dispersion relation
in $\Omega, Q$ plane. Inspecting the dispersion relation at small
$\alpha^2$ one can  observe that $f>0$ corresponds to the upper
dispersion curve and $f < 0$ to the lower one. The velocity inside
the grating is $v = 2(1-f^2)/(1+f^2)$ and equals to zero at the
edges of the gap $f = \pm 1$.

 From Eqs. (\ref{cm2}) and (\ref{cm3}) one can again see a defocusing
role of the strong nonlinearity management. We find that the
effect of nonlinearity is cancelled if $|f|=1$ and  the density of
BEC reaches a threshold value
$$\alpha_c^2 = \frac{\gamma_0}{4\epsilon\sigma^2}.$$
Suppression of the mean-field nonlinearity in the lattice leads to
enhancement of such an effect as tunnelling between sites. The SNM
also introduces changes in the dispersion curves. Indeed, it is
well known that the focusing Kerr nonlinearity (attractive BEC) is
responsible for  appearance  of a loop beyond the critical power
\cite{KivAg,Diakonov,Niu} on the upper curve.

Effective {\it nonlinear dispersion} induced by the
nonlinearity-management (the last term in Eq.~(\ref{nm_nls})) will
increase the critical power necessary for appearance of the loop.
To find this value of critical power let us consider the value of
$f_c$ at which $Q$ becomes zero ($|f|\neq 1$). We obtain that
\begin{equation}
f_c = \frac{\alpha^2(\gamma_0 - 8\epsilon\sigma^2 \alpha^2)}{2}
\pm \sqrt{\left(\frac{\alpha^2(\gamma_0 -
8\epsilon\sigma^2\alpha^2)}{2}\right)^2 -1}.
\end{equation}

 Let us consider the case of upper curve with $f
> 0$ and an attractive condensate $\gamma_0 > 0$. Then a loop
appears on the dispersion curve if the power (BEC density)
\begin{equation}\label{crit}
    \alpha_{2}^2 < \alpha^2 < \alpha_{1}^2, \  \alpha_{1,2}^2 =
    \frac{\gamma_0}{16\epsilon\sigma^2}\left(1 \pm \sqrt{1 -
    \frac{64\epsilon\sigma^2}{\gamma_0^2}}\right).
\end{equation}

When $\sigma^2 = 0$ we have a well known result for the critical
power \cite{KivAg} $\alpha_c^2 = 2/\gamma_0.$
\begin{figure}[htbp]
\begin{center}
\includegraphics[width=8cm,angle=-90]{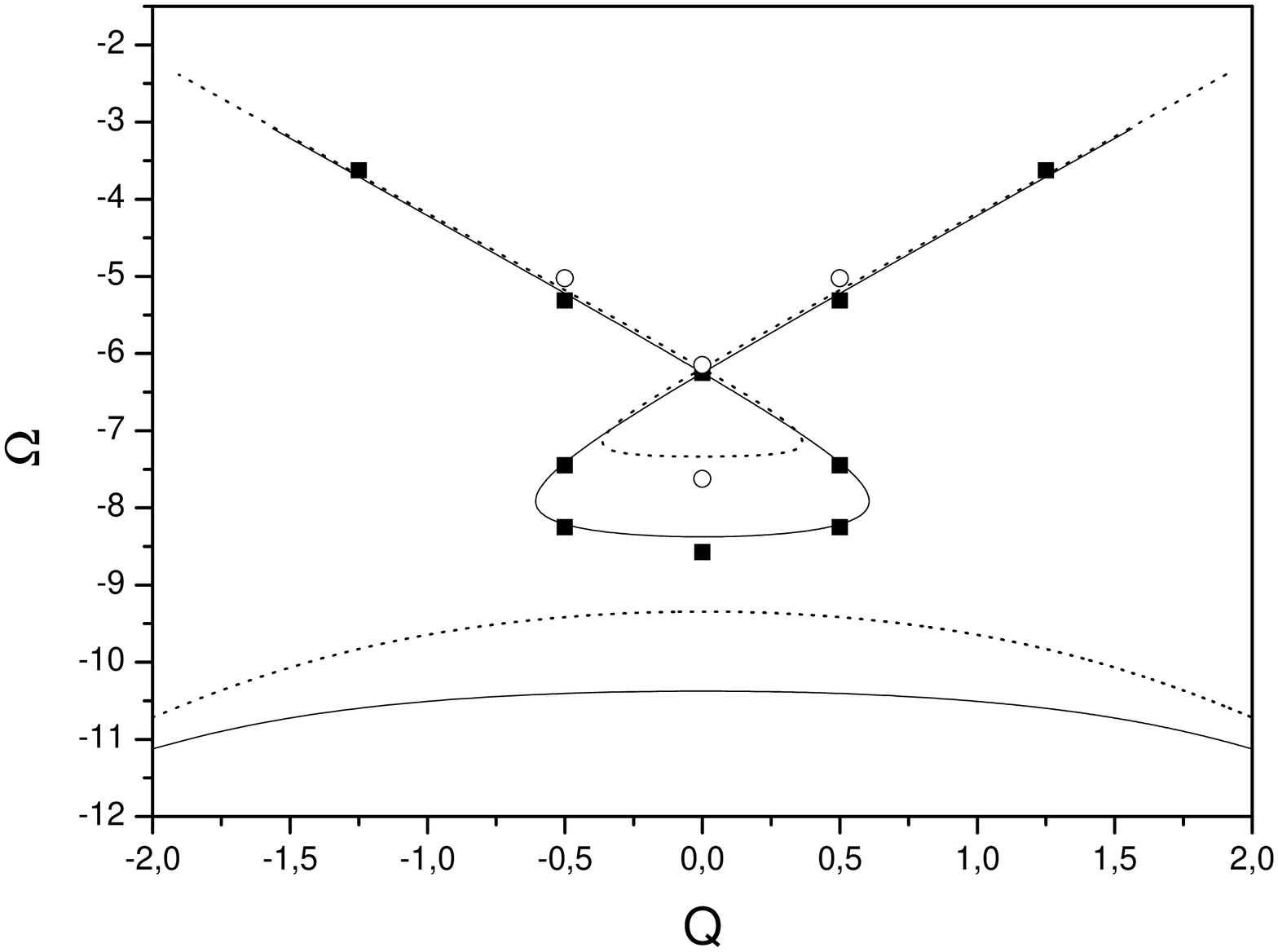}
\end{center} \hspace{1.5cm} \caption{ Loop structure in dispersion
relations when $\alpha^2 > \alpha_c^2$: solid (dotted) line and
full squares (circles) are for the case when nonlinearity
management is turned off (on). Scatter points (squares and
circles) represent data obtained from numerical simulations.
Parameters are: $\sigma^2 =0$ and $\sigma^2 =0.055 (h=4)$ with
$\alpha =2.5, \ \epsilon = 0.08, \omega = 12$.} \label{loop}
\end{figure}
Fig.~\ref{loop} depicts two branches of the dispersion relations
(\ref{cm2}) and (\ref{cm3}). The branches in the $\Omega-Q$ plane
are defined parametrically by Eq.~(\ref{cm2}) and (\ref{cm3}).
Thus, each value of $f$ defines a point in this plane. Two ranges
of values $f > 0$ and $f < 0$ define upper and lower curves
correspondingly. In our case the loop structure appears on the
upper branch when $\alpha$ is greater than the threshold value
$\alpha_c^2$. One can see from this figure that the loop decreases
with increasing of the strength of the management. It should be
noted that at the same time the band width (a distance between
upper and lower branches) at $Q = 0$ does not change.

 In the case of the defocusing Kerr nonlinearity $\gamma_0 < 0$
(repulsive BEC) one could expect formation of the loop on the
lower branch of the dispersion curve $f < 0$. But from the
condition (\ref{crit}) it follows that $\alpha^2 < 0$. So in this
case NM fully suppresses the loop formation.

 It is also of interest to investigate the loop structure in the case
$\gamma_0 = 0$.  This configuration can be realized employing the
Feshbach resonance technique. It corresponds to the case of a BEC
with the {\it effective repulsive} nonlinearity in an optical
lattice. The loop will be formed on the lower branch of the
dispersive curve $f < 0$ when the BEC density excesses the value
$$\alpha^2 > \frac{1}{2\sqrt{\epsilon}\sigma}.$$

Let us discuss the physical consequences. Existence of a loop at
the edge of the Brillouin zone reflects the superfluid character
of the BEC, since we have nonzero velocity in the Bragg reflection
condition \cite{Diakonov,Niu}. It should be noted that in a linear
system of free atoms the Bloch wave at the zone edge has zero
velocity. From this point of view a critical value of the SNM
strength exists which destroys the superfluid property of the BEC
in an optical lattice. Another possible effect is the existence of
breakdown of Bloch oscillations due to the tunnelling into the
upper band (Landau-Zeener tunnelling). The SNM is expected to
suppress this breakdown.

\subsection{Modulational instability}
To investigate MI of matter waves in an optical lattice under SNM,
perturbed plane wave solutions are taken in the form
\begin{eqnarray}\label{pw}
A & = \left(\frac{\alpha}{\sqrt{1+f^2}}+ \delta
A(X,T)\right)e^{i(QX - \Omega T)}, \\ \ B & = \left(\frac{\alpha
f}{\sqrt{1+f^2}} + \delta B(X,T)\right)e^{i(QX - \Omega T)},
\nonumber
\end{eqnarray}
where $\delta A$ and $ \delta B$ are unknown small perturbations
of CW solutions. Substituting these expressions into Eqs.
(\ref{cm1}) and (\ref{cm1a}) and using a linear approximation, we
get the system of equations for $\delta A$ and $\delta B$
\begin{eqnarray}\label{corr}
& i\delta A_T + 2i\delta A_X +  f\delta A -
    \delta B + \frac{\alpha^2}{1+f^2}[(\gamma_0 -
    \frac{16\epsilon\sigma^2\alpha^2f^2}{1+f^2}) (\delta A +
    \delta A^{\ast}) + \nonumber \\
&     2 f(\gamma_0 -
    8\epsilon\sigma^2\alpha^2)(\delta B + \delta B^{\ast})] = 0, \\
& i\delta B_T - 2i\delta B_X + \frac{1}{f}\delta B -
     \delta A + \frac{\alpha^2 f^2}{1+f^2}
    [(\gamma_0 - \frac{16\epsilon\sigma^2\alpha^2}{1+f^2})
    (\delta B + \delta B^{\ast}) + \nonumber \\
&    \frac{2}{f}(\gamma_0 -
    8\epsilon\sigma^2\alpha^2)(\delta A + \delta A^{\ast})] = 0.
\end{eqnarray}
For $f =  \pm 1$ the system coincides with the one considered by
de Sterke \cite{Sterke} with renormalized nonlinearity coefficient
$\gamma_r = \gamma_0 - 8\epsilon\sigma^2\alpha^2$. One can see
that the NM plays essential role in the MI process. When the
nonlinearity management is turned off, for the case of attractive
condensate the CW wave is unstable if the parameters follow the
upper branch of the dispersion curve. On the lower branch the
attractive BEC is modulationally stable. The repulsive condensate
is modulationally unstable on the lower branch and stable on the
upper branch.

In the case of nonlinearity-management there exists a critical
value of the management strength $\sigma^2 $, namely $\sigma_c^2 =
\gamma_0/8\epsilon\alpha^2 $. If $\sigma^2 > \sigma_c^2 $, then
the attractive condensate behaves as the repulsive and the
modulational instability regions should correspond to the above
described picture.

Looking for solutions of Eq.~(\ref{corr}) in the form
  $$\delta A(B)=C(D)\cos(qX-\omega T)+iE(F)\sin(qX-\omega T)$$
we find the dispersion relation of the form
\begin{eqnarray}\label{disp}
(\omega^2 -4q^2)^2 - 2(1-N)(\omega^2 - 4q^2)-
\frac{1}{f}(\frac{1}{f}+P)(\omega-2q)^2 - \nonumber
\\ f(f+M)(\omega + 2q)^2 =0,
\end{eqnarray}
where
\begin{eqnarray*}
M &=& \frac{2\alpha^2}{1+f^2}(\gamma_0 -
\frac{16\epsilon\sigma^2\alpha^2 f^2}{1+f^2}), \
N=\frac{4f\alpha^2}{1+f^2}(\gamma_0 - 8\epsilon\sigma^2\alpha^2),\\
P&=&\frac{2\alpha^2 f^2}{1+f^2}(\gamma_0
-\frac{16\epsilon\sigma^2\alpha^2}{1+f^2}).
\end{eqnarray*}

 Analytical results can be obtained for the particular case
  $|f|=1$, corresponding to the edges of the gap.
We come to the equation for the frequency $\omega$
\begin{eqnarray}
\omega^2 = 4q^2 + 2 - \tilde{G} \pm \sqrt{16q^2 \left(1 +
\widetilde{G} \right) + \left(2 - \widetilde{G}\right)^2 }, \\
\widetilde{G}=G/f, \ f=\pm 1. \nonumber
\end{eqnarray}
Evidently, this equation coincides with the one obtained in
Ref.~\cite{Sterke} where the parameter $G$ is renormalized as $G =
(\gamma_0 - 8\epsilon\sigma^2 \alpha^2 )\alpha^2$.

Let us analyze the condition of MI for different sets of
parameters.

\begin{figure}[htbp]
\begin{center}
\includegraphics[width=5cm]{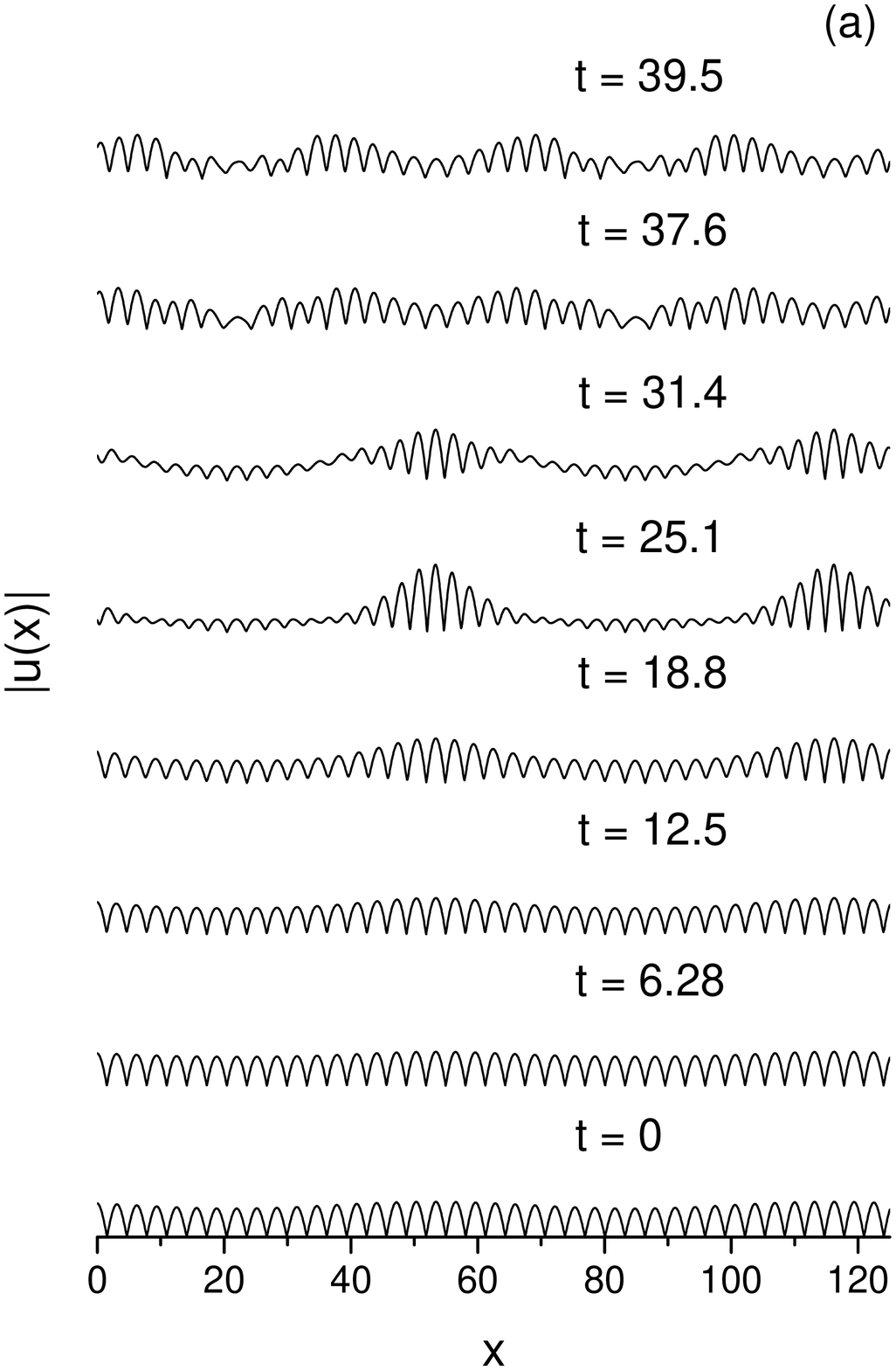}
\hspace{0.5cm}
\includegraphics[width=5cm]{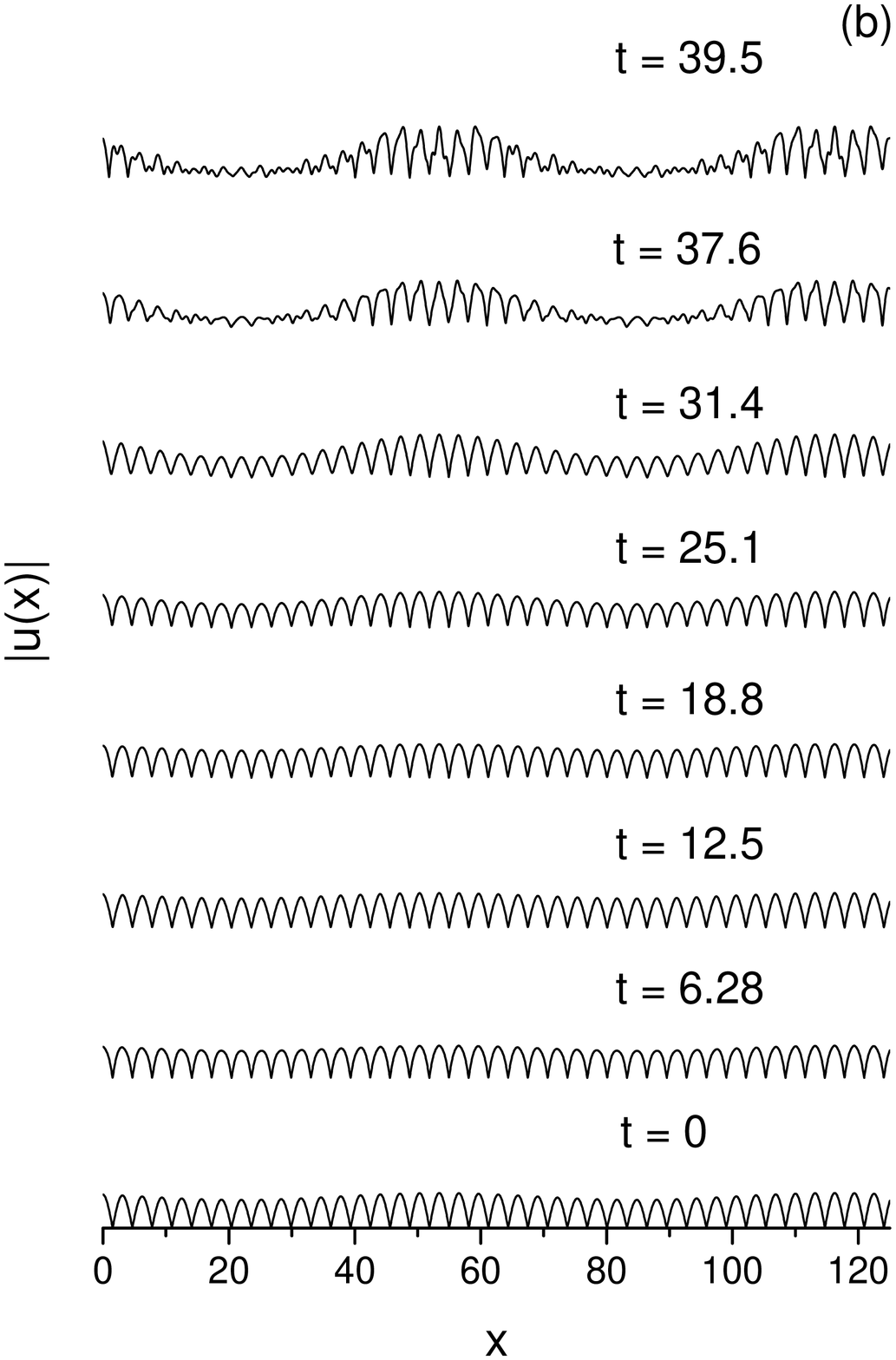}
\end{center} \hspace{1.5cm} \caption{ Evolution of small
spatial periodic perturbations when the parameters are in the
region of instability with $\gamma_0 = 1$ and $f = 1$ (upper
branch of the dispersion relations (\ref{cm2}) and  (\ref{cm3})).
Plot (a) depicts the field profiles $|u(x)|$ at different times
when the nonlinearity management is turned off and  $\sigma^2 =
0$. Plot (b) depicts the case when the nonlinearity management is
turned on and $\sigma^2 = 0.125 (h = 5)$. Other parameters are $
\alpha = 0.8, \ Q = 0, \ q = 0.5, \ \omega = 10$. Initial
amplitude of modulations is taken to be $0.05$.} \label{supgain}
\end{figure}

1. The top of the band gap $f = 1$, $\sigma^2 < \sigma_c^2$
($\widetilde G>0$). The wave is unstable if the wavenumber of
modulations is in the interval $-\sqrt{3G/2} < q < \sqrt{3G/2}$.
The maximal MI gain occurs at the wavenumber
\begin{equation}
q_m = \sqrt{3G\frac{4 + G}{16(1 + G)}}.
\end{equation}
Results of numerical simulations of the Gross-Pitaevskii equation
(\ref{nls}) for evolution of the nonlinear plane wave modulations
is shown in Fig.~\ref{supgain}. The emergence of a train of gap
solitons is observed. The reduction  of the MI gain when the SNM
is applied can be noted.

2. The bottom of the band gap $f = -1$, $\sigma^2 < \sigma_c^2$
($\widetilde G<0$). The condensate becomes unstable if $G>1$ and
the wavenumber satisfies the inequality
\begin{equation}
|q| > \frac{2 + G}{4}\sqrt{\frac{1}{G - 1}}.
\end{equation}

3. In the case $\sigma^2 > \sigma_c^2$, an {\it attractive}
condensate behaves like the {\it repulsive} condensate under the
strong nonlinearity management. We can expect modulational
instability in the case of $f=-1$, corresponding to the negative
effective mass. In this case the condensate is unstable in the
region of modulations with the wave numbers $q^2 < 3 |G|/2.$

\begin{figure}[htbp]
\begin{center}
\includegraphics[width=5cm]{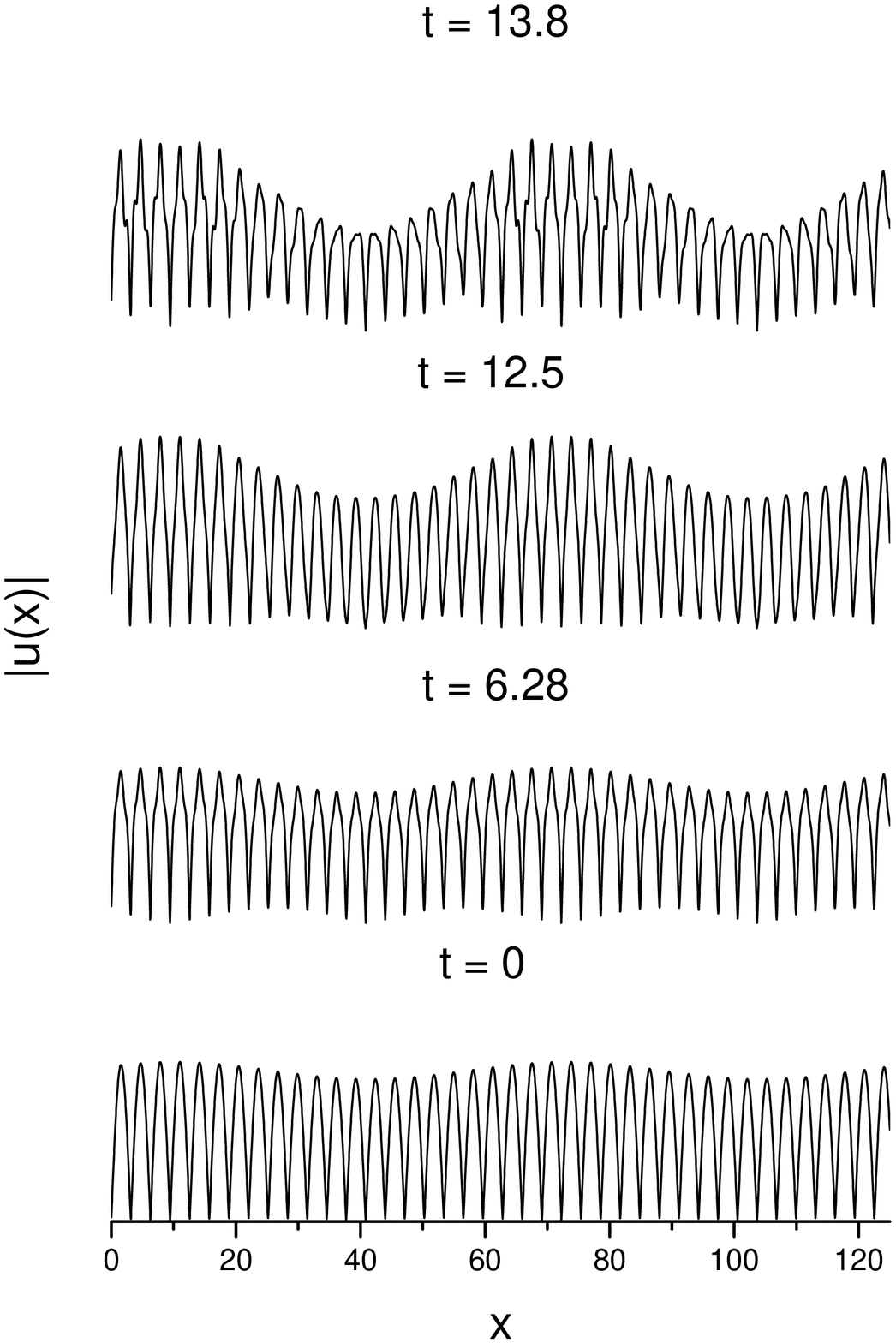}
\end{center} \hspace{1.5cm} \caption{ Evolution of small
spatial periodic perturbations when parameters are in the region
of instability with $\gamma_0 = 0$  and $f = -1$ (lower branch of
the dispersion relations (\ref{cm2}) and (\ref{cm3})). The plot
depicts the field profiles $|u(x)|$ at different times when the
strength of the nonlinearity-management is $\sigma^2 = 0.08 (h =
4)$. Other parameters are $\alpha = 1.5, \  Q = 0, \  q = 0.5, \
\omega = 10$. Initial amplitude of modulations is   $0.05$.}
\label{enhgain}
\end{figure}

Let us consider separately the case $\gamma_0 = 0, \ G = -
8\epsilon\sigma^2\alpha^4$.   As it was shown in \cite{Porter1},
near the upper edge of the gap, the gap soliton is  the solution
of the focusing quintic NLSE, while near the bottom of the gap it
is a solution of the defocusing quintic NLSE.  Fig.~\ref{enhgain}
depicts the formation of a gap soliton train under strong
nonlinearity management. It should be noted that when $\gamma_0 =
\sigma^2 = 0$, the soliton does not form.

For $f = -1$ the instability region is $|q| <
\sqrt{12\epsilon\sigma^2\alpha^2/2}.$
  For  $|f|\neq 1$  we can perform analytical consideration for the case
 of vanishing wave numbers of modulations  $q=0$. Then in the ordinary optical lattice
 the  gain of MI turns to be finite and for the MI  in the normal dispersion region there exists
 a  threshold in the power. In the case of the action of a SNM we find from Eq.~(\ref{disp})
that the  instability occurs if
 \begin{equation}
 \frac{(1+f^2)^2}{f^2} - \frac{4f(\gamma_0
 -8\epsilon\sigma^2\alpha^4)}{1+f^2} < 0.
 \end{equation}

 For example, if $\gamma_0 > 0$ and $f > 0$ the MI is possible
 only if $\alpha_2^2 < \alpha^2 < \alpha_1^2,$ where
 $$\alpha_{1,2}^2 = \frac{\gamma_0}{16\epsilon\sigma^2}\left(1 \pm \sqrt{1 -
 \frac{8(1+f^2)^3\epsilon\sigma^2}{f^3\gamma_0^2}}\right).$$
 The MI interval on  $\alpha^2$ for $f<0$ can be obtained analogously.

\section{Gap soliton}
Following the  works \cite{AbdAbdGal,Porter1} let us study the
properties of a gap soliton. The solution is sought in the form $A
= a(X)\exp(-i\bar{\Omega} T),\ B = b(X)\exp(-i\bar{\Omega} T), \ a
= b^{\ast},\ a = \sqrt{Q(X)}\exp(-i\theta(X)/2).$  The set of
equations for $Q(X),\ \theta(X)$ is
\begin{eqnarray}
Q_{X} =  Q\sin(\theta ),\\
\theta_{X}=-\bar{\Omega}+\cos(\theta)-3\gamma_0
Q+24\epsilon\sigma^2 Q^2.
\end{eqnarray}
The first integral of this set is
$$
     E = -\bar{\Omega} Q + Q\cos(\theta)-
    \frac{3}{2}\gamma_0 Q^2 + 8\epsilon\sigma^2 Q^3.
$$
Inside the gap $-1 \leq \Omega \leq 1$. The solution for $\gamma_0
\neq 0 $ is difficult to be derived  in an explicit form.
 What we
can calculate is the peak value of gap soliton amplitude,  the
quantity, which is of interest for the experiment. For the soliton
peak the condition $Q_{x} = 0$ is valid. Taking into account that
for bright soliton solution $E =0$, we obtain the following
equation for the peak value of the soliton amplitude
$$
\pm 1 = \bar{\Omega} + \frac{3}{2}\gamma_0 Q - 8\epsilon\sigma^2
Q^2,
$$
where the signs $\pm $ correspond to $\theta =0$ and $\theta =
\pi$ respectively. Peak values, corresponding to the bright
soliton solutions, are
\begin{equation}\label{lgs}
Q = \frac{3}{32\epsilon\sigma^2}\gamma_0 \left[1 - \sqrt{1 +
\frac{128\epsilon\sigma^2(\bar{\Omega} \mp
1)}{9\gamma_0^2}}\right].
\end{equation}
It should be noted that when $\gamma_0 =0 $ we get  $Q = \sqrt{(
\bar{\Omega}\mp 1)/8\epsilon\sigma^2}$, that coincides with the
value obtained in \cite{Porter1}.
 Existence of two families of gap solitons has similarity with
the ones observed in the cubic-quintic NLSE with a periodic
potential \cite{Atai}. From (\ref{lgs}) we obtain the restriction
$$\sigma^2 < \frac{9\gamma_0^2}{128\epsilon(1 - \bar{\Omega})}.$$
For the estimations of the experiment with $\epsilon = 0.2,
\bar{\Omega} = 0.6$, we obtain the restriction $\sigma^2 < 0.6$.
The defocusing role of the nonlinearity management leads to the
possibility of increasing the number of atoms in the bright gap
soliton in comparison with a standard gap soliton. The low
nonlinearity requires the larger number of atoms to support
soliton solution. Taking $\sigma^2 =0$ in the low amplitude
solution, we obtain for the peak amplitude the value $Q_0 = 2(1
-\bar{\Omega})/3,$ that reproduce the standard result for a gap
soliton \cite{AW}. Expanding the solution (\ref{lgs}) in series,
we obtain
$$
Q \approx Q_{0} + \frac{64\epsilon(1 - \bar{\Omega})^2
\sigma^2}{27\gamma_0^3}.
$$
The number of atoms in the gap soliton is enhanced and the
enhancement factor is proportional to the nonlinearity map
strength $\sigma^2$. For typical values of parameters $V = 0.6
E_{R} \ (\epsilon = 0.3),\ h = 3.16\omega , \ f= 33, \ \omega =
10\omega_R \ (\sigma^2 = h^2/2\omega^2 = 5),\ \bar{\Omega} = -1, \
\gamma_0 = 1$ we obtain double enhancement in the number of atoms.
It means that for the experiment with $^{87}$Rb \cite{Oberthaler}
the number of atoms in a gap soliton  ($N \sim 600$) can be
increased by the nonlinearity management up to $N \sim 1200$. The
increasing of number of atoms in the discrete breather of discrete
nonlinear Schr\"odinger equation under weak nonlinearity
management has been observed in numerical simulations \cite{ATMK}.

\section{Numerical simulations}
In numerical simulations we proceed from the governing
Gross-Pitaevskii equation (\ref{nls}). The problem is discretized
in a standard way with the time step $\Delta t$ and spatial step
$\Delta x$ so that terms $u_j^k$ approximate $u(j\Delta x,k\Delta
t)$. More specifically, in the approximation of Eq.~(\ref{nls}) we
have used the following implicit Crank-Nicholson-type scheme of
second order accuracy in space and first order accuracy in time
\begin{eqnarray}\label{discrGP}
\frac{i(u^{k+1}_{j} - u^{k}_{j})}{\Delta t} & = &
-\frac{1}{2\Delta x^{2}}\left[ (u^{k+1}_{j-1} - 2u^{k+1}_{j} +
u^{k+1}_{j+1}) +
(u^{k}_{j-1} - 2u^{k}_{j} + u^{k}_{j+1})\right] + \nonumber \\
& & \epsilon \cos(2x_j)(u^{k}_{j} + u^{k+1}_{j}) -
\frac{1}{2}\gamma(t_k)|u^{k}_{j}|^2 (u^{k}_{j} + u^{k+1}_{j}),
\end{eqnarray}
where the strong nonlinearity management factor $\gamma(t)$ is
defined by Eq.~(\ref{nls}), $x_k = j\Delta x$ and $t_k = k\Delta
t$. In our calculations the second term in Eq.~(\ref{nls}) is
chosen as $\gamma_{1} = h\sin(\omega t)$. For this case $\sigma^2
= h^2/(2\omega^2)$.

Since our problem deals with nonlinear plane waves, periodic
boundary conditions are imposed on the governing Eq.~(\ref{nls}).
Eq.~(\ref{discrGP}) together with the boundary condition
$u_0^{k+1} = u_{N}^{k+1}$ form a quasi tridiagonal set of
equations for unknown $u^{k+1}_{j}, \ [j = 0,1,2 ... N]$ in a
lattice of $N+1$ points. The length of the lattice $L$ is
determined by the period of the periodic potential and value of
the wave number for which the solution is sought. The set of these
algebraic equations is solved by the modified vectorial sweep
method. In actual calculations the typical space step $\Delta x$
ranged from 0.01 to 0.005 and time step $\Delta t$ from 0.005 to
0.001.

In calculations, the initial wave packet is constructed in the
following way. At first slow component of the solution $w(x,t=0)$
is taken in the form of Eq.~(\ref{perpw}) or Eq.~(\ref{pwloop})
with Eq.~(\ref{pw}), depending on the problem we consider. Then
leaving only first term in Eq.~(\ref{exp}) and making use of
transformation Eq.~(\ref{trans}) we obtain actual initial wave
function $u(x,t=0)$ used in computations.

In simulation of the loop structure (see Fig.~\ref{loop}) and
constructing initial wave function, the position on the loop for
given value of the wave number $Q$ is determined by choosing
necessary value of the parameter $f$, which, in turn, is
determined from the dispersion relation Eq.~(\ref{cm2}).

\section{Conclusion}
We have investigated the modulational instability and gap soliton
formation in the media with Kerr nonlinearity and periodic
potential. Such systems appear in the nonlinear optical media with
Bragg grating and  Bose-Einstein condensates in optical lattices
under time-dependent Feshbach resonance management. We considered
the case of strong management and showed that in the case of
homogeneous Kerr media under NM the gain of MI is strongly
suppressed, that explains the defocusing role of the NM and thus
the stabilization of 2D and 3D attractive BEC by this method. We
have studied the nonlinear dispersion relation in the case of NM
and showed that the loop structure is essentially modified by the
NM. The critical value of the strength of the NM is shown to exist
in the MI regions. In the case of attractive condensate it means
that above the threshold an attractive BEC behaves as repulsive.
The NM leads to a new effect of enhancement of the number of atoms
in the bright gap soliton. The enhancement factor is proportional
to the strength of the management $\sigma^2$. We confirmed the
predictions based on the analysis of the averaged GP equation by
direct numerical simulations of the 1D GP equation.

\section{Acknowledgements}
F.Kh.A. is grateful to IFT UNESP for the hospitality and to FAPESP
for a partial support of this work. The authors also acknowledge
B.B. Baizakov and E.N. Tsoy for useful discussions.

\end{document}